\documentclass[english,twoculumns,preprint]{revtex4}
\usepackage[latin9]{inputenc}
\usepackage{amstext,bm}
\usepackage{graphicx}
\usepackage{xcolor}
\usepackage{color}
\usepackage{amsmath}
\usepackage{babel}
\usepackage[titletoc]{appendix}
\usepackage{hyperref}
\usepackage{float}

\newcommand{\nabi}{Na$_3$Bi }

\newcommand{\bise}{Bi$_2$Se$_3$ }

\begin{document}

\title{Topological Nodal Line Semimetal and Dirac Semimetal State in Antiperovskite Cu$_3$PdN}

\author{Rui Yu$^{1}$}
\author{Hongming Weng$^{2,3}$}
\email{hmweng@iphy.ac.cn}
\author{Zhong Fang$^{2,3}$}
\author{Xi Dai$^{2,3}$}
\author{Xiao Hu$^{1}$}
\email{Hu.Xiao@nims.go.jp}


\affiliation{ $^{1}$ International Center for Materials Nanoarchitectornics (WPI-MANA) National Institute for Materials Science, Tsukuba 305-0044, Japan}
\affiliation{$^{2}$
Beijing National Laboratory for Condensed Matter Physics, and Institute of
Physics,Chinese Academy of Sciences, Beijing 100190, China}

\affiliation{$^{3}$
Collaborative Innovation Center of Quantum Matter, Beijing 100190, China}

\begin{abstract}
Based on first-principles calculation and effective model analysis, we propose that the cubic antiperovskite material Cu$_3$PdN can host a three-dimensional (3D) topological
nodal line semimetal state when spin-orbit coupling (SOC) is ignored, which is protected by coexistence of time-reversal and inversion symmetry.
There are three nodal line circles in total due to the cubic symmetry.
``Drumhead"-like surface flat bands are also derived.
 When SOC is included, each nodal line evolves into a pair of stable 3D Dirac points as protected by C$_4$ crystal symmetry.
This is remarkably distinguished from the Dirac semimetals known so far, such as Na$_3$Bi and Cd$_3$As$_2$, both having only one pair of Dirac points.
Once C$_4$ symmetry is broken, the Dirac points are gapped and the system becomes a strong topological insulator with (1;111) Z$_2$ indices.

\end{abstract}

\date{\today}


\maketitle

\noindent\textit{Introduction ---}
Band topology in condensed matters has attracted broad interests in recent years. It was prospered by the discovery of two-dimensional (2D) and three-dimensional (3D)
topological insulators (TIs)
\cite{RMP_Kane_2010,RMP_Qi_2011}.
These materials exhibit a bulk energy gap between the valence and conduction bands, similarly to normal insulators, but with gapless boundary states that are protected by the band topology of bulk states.
{Topological classification has also been proposed for 3D metals \cite{Horava_2005PRL}.}
The topological invariant, so called Fermi Chern number,
{can be}
defined on a closed 2D manifold, such as the Fermi surface, in 3D momentum space~\cite{Horava_2005PRL, Wang:2012ds, MRS_weng:9383312}.
This is essentially the same as Chern number defined in the whole 2D Brillouin zone (BZ) for insulators.
Up to now, three types of nontrivial topological metals have been proposed.
They are Weyl semimetal \cite{Nielsen:1983ce,WanXianGang:2011PRL,XuGang:2011PRL,Balents_Weyl:2011vy}, Dirac semimetal \cite{Wang:2012ds,Wang:2013is} and nodal line semimetal (NLS) \cite{Burkov:2011ega, WengHM_2014_Graphene_Networks}.
All of them have band crossing points due to band inversion~\cite{Yang_BohmJung_NC_2014}.
For Weyl and Dirac semimetals, crossing points locate at different positions which compose Fermi surfaces.
%
For the NLS, the crossing points around the Fermi level form a closed {loop}.
The breakthrough in topological semimetal research happened in material realization of Dirac semimetal state in \nabi and Cd$_3$As$_2$, which were first
predicted theoretically ~\cite{Wang:2012ds,Wang:2013is} and then confirmed by several experiments \cite{Liu:2014bf,Liu:2014hr,Neupane:2014kc,Jeon:2014ii}.
Starting from Dirac semimetal, one can obtain Weyl semimetal by breaking either
time-reversal~\cite{WanXianGang:2011PRL,XuGang:2011PRL,TRSB_Weyl_Zyuzin_PhysRevB} or inversion symmetry~\cite{inversion_Wyel_Halasz_prb,inversion_Wyel_Hirayama:2014uf,inversion_Wyel_Vanderbilt,inversion_Wyel_Murakami:2007ig}.
Among them, the prediction of Weyl semimetal state in nonmagnetic and noncentrosymmetric TaAs family~\cite{Weng_TaAs_PhysRevX.5.011029,TaAs_Huang:2015uu} have been verified
by experiments~\cite{TaAs_exp_Lv:2015ws,Weyl_exp_lv_observation_2015,weyl_exp_Genfu_Chen,TaAs_exp_Xu:2015tb,TaAs_exp_Shekhar:2015wj,weyl_exp_NbAs}.

Existence of NLS has been proposed in Bernal graphite~\cite{GP_Mikitik_2006PRB, GP_Mikitik_2008LTP} and other all-carbon allotropes, including hyper-honeycomb lattices~\cite{nodal_line_2014arXiv_Hyper_Honeycomb_Lattices} and Mackay-Terrones crystal (MTC)~\cite{WengHM_2014_Graphene_Networks}. In Ref.~\onlinecite{WengHM_2014_Graphene_Networks}, it was discussed that when spin-orbit coupling (SOC) is neglected and band inversion happens, the coexistence of time-reversal and inversion symmetry  protects NLS in 3D momentum space.
It is heuristic that compounds with light element can also exhibit nontrivial topology, which is totally different from the common wisdom that strong SOC in heavy-element compounds is crucial for topological quantum states.
The NLS state was also proposed in photonic systems with time-reversal and inversion symmetry~\cite{LuLing_weyl_photonic}.
It was also shown that mirror symmetry, instead of inversion symmetry, together with time-reversal symmetry protects NLS when SOC is neglected~\cite{Schnyder_2014PRB, FanZhang_2014PRL}. TaAs~\cite{Weng_TaAs_PhysRevX.5.011029}, Ca$_3$P$_2$~\cite{Ca3P2_Xie:2015tq} and LaX (X=N, P, As, Sb, Bi)~\cite{Fuliang_2015arXiv150403492Z} {belong to this type.}

In the present work, based on first-principles calculations and effective model analysis, we demonstrate that the antiperovskite Cu$_3$PdN is a new candidate for realizing  NLS
and ``drumhead"-like surface flat bands~\cite{Ryu_2002PRL, Heikkila_2011JETP1,Heikkila_2011JETP2,Heikkila_2015arXiv03277,Volovik_2011,WengHM_2014_Graphene_Networks}, {which may  open an important route to achieving high-temperature superconductivity~\cite{Heikkila_2011PRB,Volovik_2014arXiv,Heikkila_2015arXiv}.}
Strong SOC will drive NLS in Cu$_3$PdN into Dirac semimetal state with three pairs of Dirac points, leading to exotic surface Fermi arcs which
can be observed on various surfaces of this material. This is very unique since Dirac semimetals known so far, \nabi and Cd$_3$As$_2$, have only one pair of Dirac points. When the C$_4$ crystal symmetry in Cu$_3$PdN is broken,
the Dirac points are gapped and the system becomes a strong TI with Z$_2$ indices (1;111). It is well known that the
 cooperative interactions among lattice, charge, and spin degrees of freedom make antiperovskites exhibit a wide range of interesting physical properties, such as superconductivity \cite{perovskite_SC}, giant magnetoresistance \cite{perovskite_magnetoresistance}, negative thermal expansion \cite{perovskite_negative_thermal}, and magnetocaloric effect \cite{perovskite_magnetocaloric} . Our prediction of 3D NLS and Dirac semimetal state in antiperovskites provides a promising platform for manipulating these exotic properties in presence of nontrivial topology.

\begin{figure}[t]
\begin{centering}
\includegraphics[width=1\columnwidth]{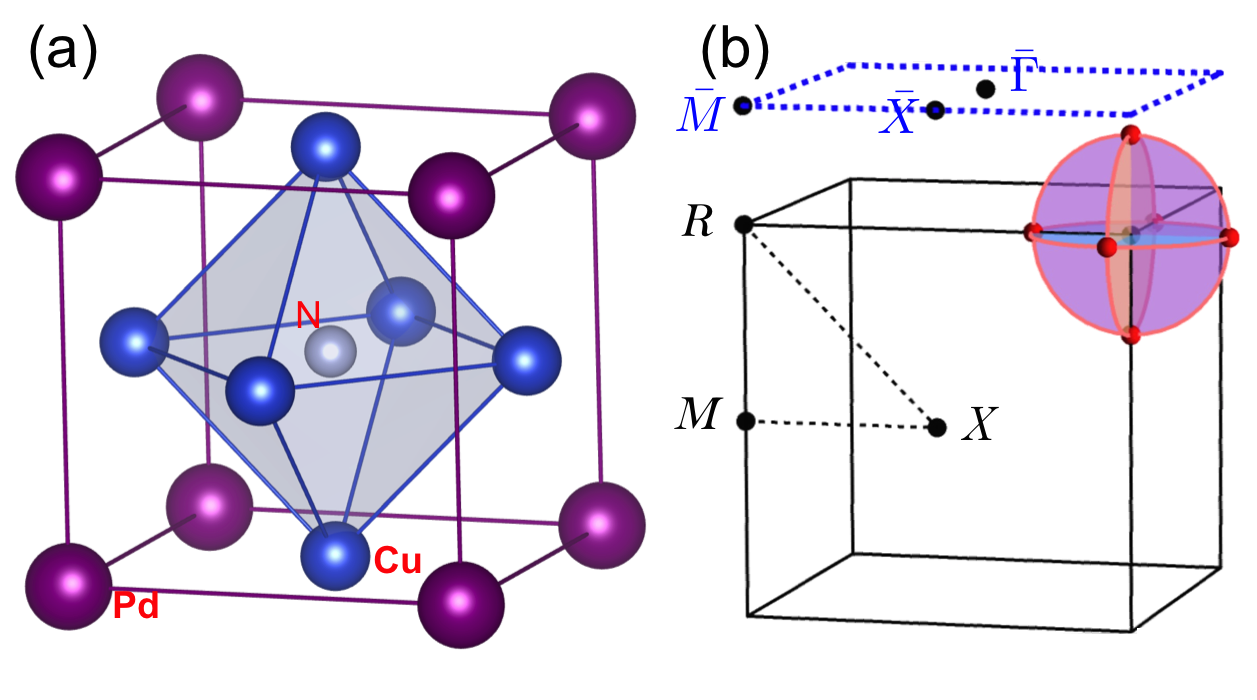}
\par\end{centering}

\protect\caption{\label{fig:crystal_BZ}(Color online)
(a) Crystal structure of antiperovskite Cu$_3$PdN with Pm$\bar{3}$m (No.221) symmetry. Nitrogen atom is at the center of the cube and is surrounded by octahedral Cu atoms. Pd is located at the corner of the cube.
(b) Bulk and projected (001) surface Brillouin zone. The three nodal line rings (orange color)  and three pairs of Dirac points (red points) without and with SOC included, respectively, are schematically shown.}
\end{figure}

\vspace{3mm}
\noindent\textit{Crystal structure and methodology ---}
Perovskite has a formula ABX$_3$, where A and B are cations and X is an anion. The antiperovskite parallels to perovskite but switching the position of anion and cation, namely in ABX$_3$ X stands for an electro-positive cation, while A for an anion as shown in Fig. \ref{fig:crystal_BZ}(a) for Cu$_3$PdN.
Here we perform density functional calculations by using the Vienna {\it ab initio} simulation package \cite{VASP} with generalized gradient approximation~\cite{PBE} and the projector augmented-wave method~\cite{PAW_Blochl:1994uk}.
The surface band structures are calculated in a tight-binding scheme based on the maximally localized Wannier functions (MLWF) \cite{Vanderbilt_RMP}, which are projected from the bulk Bloch wave functions.


\begin{figure}[t]
\begin{centering}
\includegraphics[width=1\columnwidth]{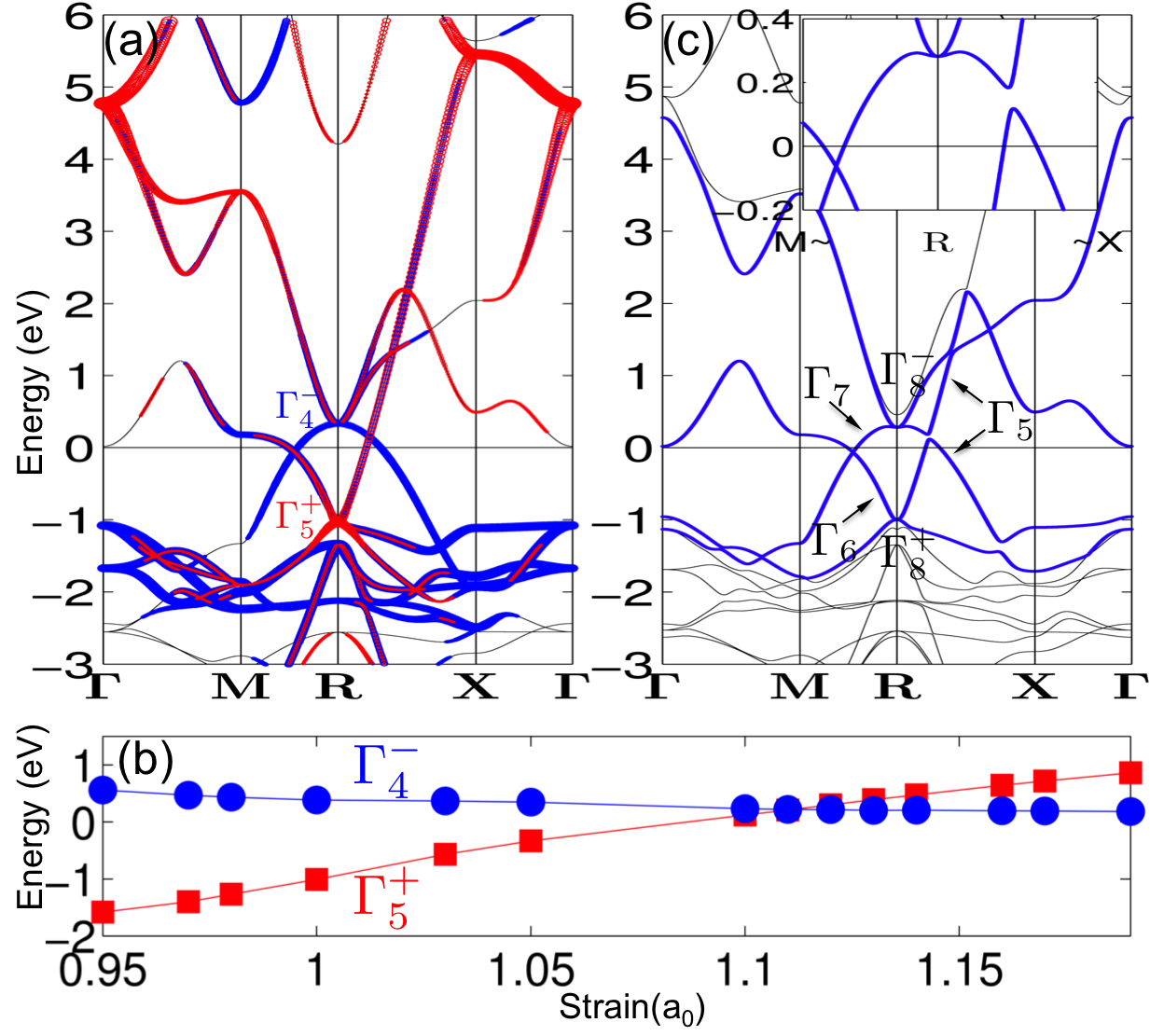}
\par\end{centering}
\protect\caption{\label{fig:bnd}(Color online)
(a) Electronic band structure without SOC, where the component of Pd-5$p$ (4$d$) orbitals is proportional to the width of the red (blue) curves. Band inversion between $p$ and $d$ orbits happens at R point.
(b) Evolution of energy levels of $\Gamma_5^+$ and $\Gamma_4^-$ at R under hydrostatic pressure. Band inversion happens when $a<1.11a_0$.
(c) Electronic band structure with SOC included. A small gap is opened in the R-X direction while the Dirac point in the R-M direction is stable and protected by crystal symmetry C$_4$ rotation.
}
\end{figure}

\vspace{3mm}
\noindent\textit{Electronic structures ---}
The band structure of Cu$_3$PdN is shown in Fig.~\ref{fig:bnd}.
The fatted bands suggest that the valence and conduction bands are dominated by Pd-4$d$ (blue) and Pd-5$p$ (red) states. They also indicate band inversion at R point, where the Pd-5$p$ is lower than
Pd-4$d$ by about 1.5 eV. To overcome possible overestimation of band inversion~\cite{Zunger_PhysRevB}, we employ hybrid density functionals \cite{HSE_2003} to confirm the existence of band inversion, while the energy gap at R point is slightly reduced to 1.1 eV.

Without SOC the occupied and unoccupied low energy bands are triply degenerate at R point (six-fold degenerate if spin is considered).
These states belong to the 3D irreducible representations $\Gamma_4^-$ and $\Gamma_5^+$ of $O_h$ group at R point, respectively.
We emphasize that, unlike the situation in typical TI such as the \bise family compounds \cite{Zhang:2009ks}, the band inversion in Cu$_3$PdN is not due to SOC.
To illustrate the band inversion process in Cu$_3$PdN explicitly, we calculated energy levels of $\Gamma_5^+$ and $\Gamma_4^-$ bands at R point for Cu$_3$PdN under different hydrostatic strains.
As presented in Fig. \ref{fig:bnd}(b), the band inversion happens at $a=1.11a_0$ and the inversion energy increases as further compressing the lattice.
The intriguing point of the Cu$_3$PdN band structure without SOC is that the band crossings due to the band inversion form a nodal line circle because of the  coexistence of time reversal and inversion symmetry as addressed in Ref. \cite{WengHM_2014_Graphene_Networks}.

The protection of a nodal line can be inferred from the topological number of the form~\cite{Ryu_2002PRL,Heikkila_2011JETP1,Heikkila_2011JETP2,Heikkila_2015arXiv03277,Volovik_2011}
\begin{equation}
\gamma=\oint_{C}\mathcal{A}
(\mathbf{k})\cdot d\mathbf{k},\label{eq:topoNum}
\end{equation}
where $\mathcal{A}(\mathbf{k})$ is the Berry connection of the occupied states,
$C$ is a closed loop in the momentum space. If $C$ is pierced by the nodal line, one has $\gamma$=$\pi$, otherwise $\gamma$=0.
Below we prove the existence of nodal line from the argument of codimension.
A general 2$\times$2 Hamiltonian is enough for describing the two crossing  bands,
\begin{equation}
H(\mathbf{k})=g_{x}(\mathbf{k})\sigma_{x}+g_{y}(\mathbf{k})\sigma_{y}+g_{z}(\mathbf{k})\sigma_{z},\label{eq:hk}
\end{equation}
where $\mathbf{k=}(k_{x},k_{y},k_{z})$, $g_{x,y,z}(\mathbf{k})$ are real functions and
$\sigma_{x,y,z}$ are Pauli matrices characterizing the space of the two
crossing bands, which are mainly from $p_{z}$ and $d_{xy}$ orbitals of Pd in Cu$_3$PdN.
The coexistence of time-reversal and inversion symmetry leads to $g_{y}=0$,
$g_{x}$ and $g_{z}$ being odd and even function of ${\bf k}$~\cite{WengHM_2014_Graphene_Networks}.
Up to the lowest order of ${\bf k}$, $g_x(\bf k)$ and $g_z(\bf k)$ are given as
\begin{align}
g_{x}(\mathbf{k}) & =\gamma k_{x}k_{y}k_{z},\label{eq:gy}\\
g_{z}(\mathbf{k}) & =M-B(k_{x}^{2}+k_{y}^{2}+k_{z}^{2}),\label{eq:gz}
\end{align}
considering the crystal symmetry at R point.
The eigenvalues of Eq.~(\ref{eq:hk}) are $E(\mathbf{k})=\pm\sqrt{g_{x}^{2}(\mathbf{k})+g_{z}^{2}(\mathbf{k})}$.
Nodal lines appear when $g_{x}(\mathbf{k})=0$ and $g_{z}(\mathbf{k})=0$.
It is easy to check that $g_{z}(\mathbf{k})=0$ can be satisfied only in the case of $MB>0$, which is nothing but the band inversion condition.
When band inversion happens, there alway exists three closed nodal lines in  momentum space as shown in Fig. \ref{fig:crystal_BZ}(b) which are the solutions of $g_x(\mathbf{k})=g_z(\mathbf{k})=0$.

When SOC is considered, the six-fold degenerate states at R are split into one four-fold and one two-fold degenerate states.
As shown in Fig. \ref{fig:bnd}(c), there are two four-fold degenerate states close to Fermi energy: the occupied one with $\Gamma_8^+$ symmetry and the unoccupied one with $\Gamma_8^-$ symmetry. Moving from R to X, the symmetry is lowered to $C_{2v}$.
The four-fold degenerate states at R are split into two-fold degenerate states along R-X direction.
The {first-principles} calculation shows that two sets of {bands} with the same $\Gamma_5$ symmetry are close to the Fermi energy and a gap {$\sim$}0.062 eV is opened at the intersection as shown in the inset of Fig.~\ref{fig:bnd}(b).
{In order to reduce this SOC splitting and thus achieving NLS,
one can replace Pd with lighter elements, such as Ag and Ni.}
Along R-M direction, the symmetry is characterized by $C_{4v}$ double group.
As indicated in {Fig.~\ref{fig:bnd}(c)}, the two sets of bands close to Fermi energy belong to $\Gamma_7$ and $\Gamma_6$ representation, respectively. They are decoupled and the crossing point on R-M path is unaffected by SOC.
They form a Dirac node near the Fermi energy as shown in Fig.~\ref{fig:bnd}(c), which
is protected by the crystal symmetry C$_4$ rotation~\cite{Wang:2013is,Wang:2012ds, TlN_ShengXL_PhysRevB}.
Once C$_4$ rotational symmetry is broken, the Dirac node will be gapped and the parities of occupied states at eight time reversal invariant momenta (TRIM) point are shown in Table.~\ref{table:parity}. The Z$_2$ indices are (1;111), indicating a strong TI.
The band structure of Cu$_3$PdN is different from that of antiperovskite Sr$_3$PbO~\cite{FuLiang_antiperovskites_PhysRevB,Ca3PbO_2012JPSJ},
{where} the band inversion happens at $\Gamma$ point and the involved bands belong to the same irreducible presentation,
which leads to anti-crossing along  $\Gamma$-X direction.

\begin{table}[ht]
\begin{centering}
\begin{tabular}{c|c|c|c|c}
\hline
TRIM points & R  & $\Gamma$  & M($\times 3$) & X($\times 3$)\tabularnewline
\hline
parity product & + & -  & - & - \tabularnewline
\hline
\end{tabular}
\par\end{centering}
\protect\caption{Parity product of occupied states at the TRIM points.
{The Z$_2$ indices are (1;111).}}
\label{table:parity}
\end{table}
\vspace{3mm}
\noindent\textit{Effective Hamiltonian for the 3D Dirac fermions ---}
{In order to better understand the band crossing and gap opening discussed above, we derive an low energy effective Hamiltonian based on the theory of invariants in a similar way as that} for Bi$_2$Se$_3$ family \cite{Zhang:2009ks}.

We first construct the model Hamiltonian along R-M in $k_z$ direction.
The symmetrical operations in this direction contain the crystalline $C_{4v}$ symmetry and inversion with time-reversal symmetry (PT).
As discussed above, the wave functions of low-energy states along R-M direction are $\Gamma_6$ symmetry with angular momentum $j_z=\pm 1/2$ and $\Gamma_7$ with $j_z=\pm 3/2$. Therefore, the model Hamiltonian {respecting} both $C_{4v}$ and PT symmetries can be written as
\begin{equation}
H_{RM}=\left[\begin{array}{cccc}
\mathcal{M}_{1} & 0 & c_{2}k_{+} & c_{3}k_{-}^{2}+c_{4}k_{+}^{2}\\
 & \mathcal{M}{}_{1} & c_{4}k_{-}^{2}+c_{3}k_{+}^{2} & -c_{2}k_{-}\\
 &  & \mathcal{M}{}_{2} & 0\\
\dagger &  &  & \mathcal{M}{}_{2}
\end{array}\right]
\label{eq:Hc4v}
\end{equation}
up to the second order of ${\bf k}$ in basis of
$\{
|j_z=\frac{1}{2}\rangle_p,
|j_z=-\frac{1}{2}\rangle_p,
|j_z=\frac{3}{2}\rangle_d,
|j_z=-\frac{3}{2}\rangle_d\}$,
where $\mathcal{M}_{1}=m_{p}+m_{11}k_{z}+m_{12}k^{2}$, $\mathcal{M}_{2}=m_{d}+m_{21}k_{z}+m_{22}k^{2}$, $k^2=k_x^2+k_y^2+k_z^2$ and $k_{\pm}=k_x\pm i k_y$.
The $C_4$ rotation symmetry requires the matrix element between $|j_z=\frac{1}{2}\rangle$
and $|j_z=-\frac{3}{2}\rangle$ to take the form of $k_{\pm}^2$ in order to conserve the
total angular moment along {$z$} direction. Due to PT {symmetry} and mirror symmetry, all parameters in Eq.~(\ref{eq:Hc4v}) can be chosen {as} real. {Their values
can be derived by fitting the dispersions to those of first-principles calculations}
\footnote{The fitted parameters are listed:
$m_p=-1.12$eV, $m_d=0.254$eV,
$m_{11}=-0.86$,
$m_{12}=33.6$,
$m_{21}=-0.40$,
$m_{22}=-7.1$,
$m_3=0.1$,
$m_{31}=0.51$,
$m_{32}=-3.1$,
$c_{1}=0.3$,
$c_{2}=-0.2$,
$c_{3}=1.2$ and
$c_{4}=0.95$, where the unit of energy is eV and the unit of length is the lattice constant.}.

On $k_z$ axis, the effective Hamiltonian is diagonal and the $|\pm 1/2\rangle$ sets and $|\pm3/2\rangle$ sets are decoupled. Since the energy of $p$ and $d$ orbitals are inverted at R point, namely $m_p<m_d$ and $m_{22}<0<m_{12}$, the $|\pm 1/2\rangle$  and $|\pm 3/2\rangle$ bands cross at
$k_{z}^{c}=[((m_{11}-m_{21})^{2}-4(m_{12}-m_{22})(m_{p}-m_{d}))^{1/2}-(m_{11}-m_{21})]/[2(m_{12}-m_{22})]$ point.

The model Hamiltonian along R-X direction can be derived in the same way with $C_{2v}$ and PT symmetries.
From the {first-principles} calculations, the {states near} Fermi energy are characterized by $\Gamma_5$ with $j_z=|\pm 1/2\rangle$.
The model Hamiltonian can be written as
\begin{equation}
H_{RX}(k)=\left[\begin{array}{cccc}
\mathcal{M}_{1} & 0 & \mathcal{M}_{3} & c_{1}k_{-}\\
 & \mathcal{M}{}_{1} & c_{1}k_{+} & -\mathcal{M}_{3}\\
 &  & \mathcal{M}{}_{2} & 0\\
\dagger &  &  & \mathcal{M}{}_{2}
\end{array}\right]\label{eq:Hc2v}
\end{equation}
up to the second order of $\bf{k}$ in basis set of
$\{
|j_z=\frac{1}{2}\rangle_p,
|j_z=-\frac{1}{2}\rangle_p,
|j_z=\frac{1}{2}\rangle_d,
|j_z=-\frac{1}{2}\rangle_d\}$.
{Here, $k_z$ is set as along R-X direction}.
The term $\mathcal{M}_3=m_{3}+m_{31}k_{z}+m_{32}k^{2}$  conserves the total angular momentum along {$z$ direction. It couples the going-down $\Gamma_5$ states and going-up $\Gamma_5$ states and opens a gap at the crossing point shown in Fig.~\ref{fig:bnd}(c).}

\begin{figure}[]
\begin{centering}
\includegraphics[width=1\columnwidth]{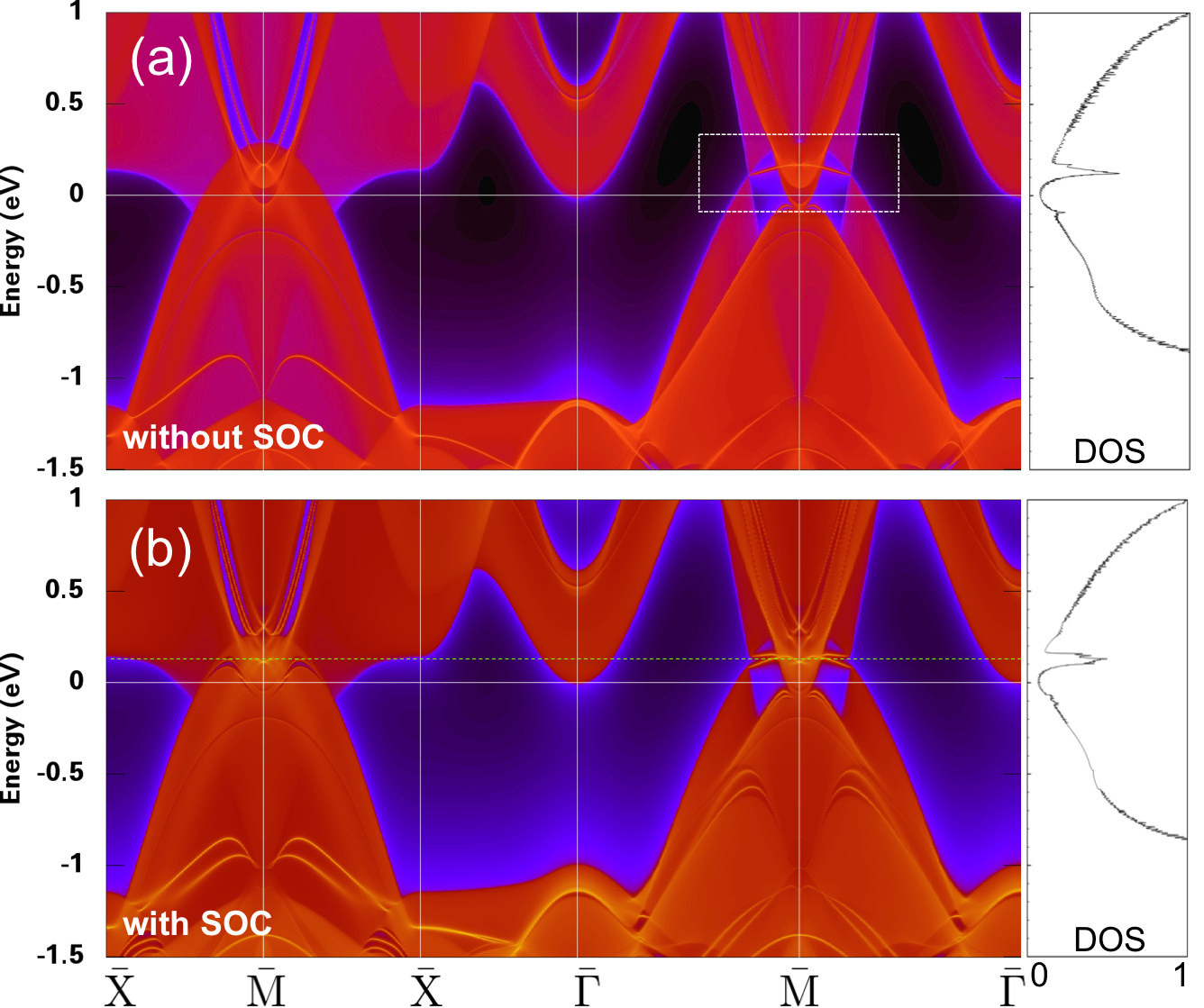}
\par\end{centering}
\protect\caption{\label{fig:sf}(Color online)
Band structures and DOS for (001) surface (a) without and  (b) with SOC .
Without SOC, the nearly flat surface bands are clearly shown in the white dashed box around $\bar{M}$ point.
}
\end{figure}

\vspace{3mm}
\noindent\textit{Surface states ---}
The band inversion and the 3D Dirac cones in Cu$_3$PdN suggest the presence of topologically nontrivial surface states.
They are calculated based on TB Hamiltonian from MLWF~\cite{Vanderbilt_RMP,MRS_weng:9383312}.
The obtained band structures and surface density of states (DOS) on semi-infinite (001) surface are presented in Fig.~\ref{fig:sf}.

Without SOC, the bulk state is the same as MTC~\cite{WengHM_2014_Graphene_Networks} and there exists surface flat bands nestled inside the projected nodal line ring on the (001) surface, namely the ``drumhead" states as shown in {Figure \ref{fig:sf}(a)}. The peak-like DOS from these nearly flat bands is also clearly shown.
The small dispersion of this ``drumhead" state comes from the fact that the nodal line ring is not necessarily on the same energy level due to {the particle-hole asymmetry}~\cite{WengHM_2014_Graphene_Networks,Burkov:2011ega,Heikkila_2015arXiv}.
{This result is consistent with the
correspondence between Dirac line in bulk and flat band at boundary as established in Ref.~\cite{Ryu_2002PRL,Volovik_2011,Heikkila_2011JETP1,Heikkila_2011JETP2}.}
{Such 2D flat bands and nearly infinite DOS are proposed as a route to achieving high-temperature superconductivity~\cite{Heikkila_2011PRB,Volovik_2014arXiv,Heikkila_2015arXiv}. }

\begin{figure}[]
\begin{centering}
\includegraphics[width=1\columnwidth]{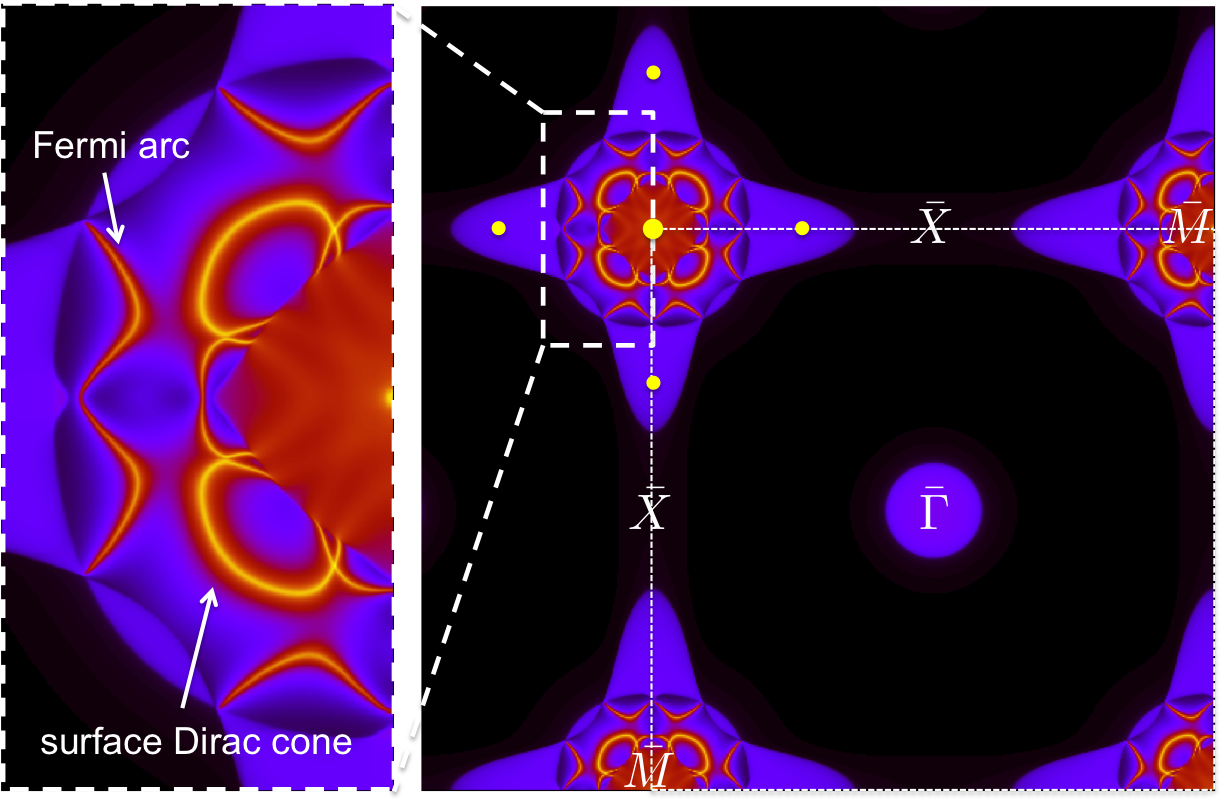}
\par\end{centering}
\protect\caption{\label{fig:SS}(Color online)
{Fermi surface of (001) surface} shown in Fig.~\ref{fig:sf}(b) with chemical potential at 0.12 eV (green dashed line in Fig. \ref{fig:sf} (b). The yellow dots are projected Dirac points. The bigger one at $\bar{M}$ means there are two Dirac points superposed there.
}
\end{figure}

In the presence of SOC, the nodal line ring will be gapped in general.
However, there is an exception.
For example, in TaAs family each ring evolves into
three pairs of Weyl nodes \cite{Weng_TaAs_PhysRevX.5.011029}. In Cu$_3$PdN, each ring is driven into one pair of Dirac nodes. The (001) surface state band structure in Fig.~\ref{fig:sf}(b)
clearly {shows}  the gapped bulk state along $\bar{\Gamma}$-$\bar{M}$ direction and the existence of  surface Dirac cone due to {topologically} nontrivial Z$_2$ indices
as seen in Na$_3$Bi~\cite{Wang:2012ds} and Cd$_3$As$_2$~\cite{Wang:2013is}.
The bulk band structure along R-X and R-M overlap each other when projected onto (001) surface along the $\bar{X}$-$\bar{M}$
path. The bulk Dirac cones are hidden {by} other bulk bands. Therefore, it is difficult to identify the detailed connection of Fermi arcs in the Fermi surface plotting as shown in Fig.~\ref{fig:SS}, though some eyebrow-like Fermi arcs can be clearly seen around these projected Dirac nodes.

\vspace{3mm}
\noindent\textit{Conclusion ---}
In summary, {we propose} that 3D topological nodal line semimetal {states} can be obtained in {a nonmagnetic and centrosymmetric system} Cu$_3$PdN. The ``drumhead"-like surface flat bands nestled {in a} projected nodal line ring have been obtained. Including SOC will drive each nodal line ring into one pair of Dirac points to host Dirac semimetal state. The surface Dirac cone and the Fermi arcs around projected Dirac cones {are observed}. The {existence of multiple pairs} of 3D Dirac points {distinguishes this system} from other Dirac semimetals with only one pair. The cubic antiperovskite structure of Cu$_3$PdN {makes it a good platform for manipulating ferromagnetism, ferroelectricity and superconductivity
realized in a broad class of materials with perovskite structure in presence of nontrivial topology}.

\vspace{3mm}
\noindent\textit{Acknowledgments ---}
{After we finalize this manuscript, we noticed that there is a similar work from Kim et al~\cite{YK_Kane_arxiv_Dirac_nodeLine_2015}.}
This work was supported by the WPI Initiative on Materials Nanoarchitectonics, and Grant-in-Aid for Scientific Research under the Innovative Area `Topological Quantum Phenomena' (no. 25103723), MEXT of Japan. HM, ZF and XD was supported by the National Natural Science Foundation of China, the 973 program of China (No. 2011CBA00108 and No. 2013CB921700), and the ``Strategic Priority Research Program (B)" of the Chinese Academy of Sciences (No. XDB07020100).


%

\end{document}